\documentclass[10pt,twocolumn,twoside]{IEEEtran}

\usepackage{graphicx}
\usepackage[cmex10]{amsmath}
\usepackage{amssymb}

\begin{document}

\title{A Receiver-Centric OFCDM Approach\\ with Subcarrier Grouping}

\author{Nikolaos~I.~Miridakis,
        Dimitrios~D.~Vergados,~\IEEEmembership{Senior Member,~IEEE} and Emmanouil~Papadakis
\thanks{N. I. Miridakis, D. D. Vergados and E. Papadakis are with the Department of Informatics, University of Piraeus, GR-185 34, Piraeus, Greece (e-mail: \{nikozm; vergados\}@unipi.gr, mpapadak@students.cs.unipi.gr).}
}

\markboth{IEEE Communications Letters}%
{A Receiver-Centric OFCDM Approach with Subcarrier Grouping}

\maketitle

\begin{abstract}
In this letter, following a cross-layer design concept, we propose a novel subcarrier grouping technique for Orthogonal Frequency and Code Division Multiplexing (OFCDM) multiuser systems. We adopt a two dimensional (2D) spreading, so as to achieve both frequency- and time-domain channel gain. Furthermore, we enable a receiver-centric approach, where the receiver rather than a potential sender controls the admission decision of the communication establishment. We study the robustness of the proposed scheme in terms of the Bit-Error-Rate (BER) and the outage probability. The derived results indicate that the proposed scheme outperforms the classical OFCDM approach. 
\end{abstract}

\begin{IEEEkeywords}
Orthogonal Frequency and Code Division Multiplexing (OFCDM), Subcarrier Grouping, Bit-Error-Rate (BER) Performance, Outage Probability, Cross-Layer Design.
\end{IEEEkeywords}

\IEEEpeerreviewmaketitle

\section{Introduction}
\IEEEPARstart{O}{rthogonal} Frequency and Code Division Multiplexing (OFCDM) was recently proposed for multiuser access in the fourth generation (4G) wireless framework \cite{ref7,ref2}. The adoption of a two dimensional (2D) spreading in OFCDM, by utilizing both frequency- and time-domain spreading, is shown to outperform the MultiCarrier Code Division Multiple Access (MC-CDMA) and the Direct Sequence MC-CDMA (MC-DS-CDMA) schemes \cite{ref1}. This is achieved because each data stream is spread simultaneously over multiple subcarriers and multiple OFCDM symbols. Hence, by exploiting both the time and the frequency diversity of the channel, a performance gain is straightforward. Furthermore, OFCDM can facilitate the power limitation issue in multiuser environments, especially in the demanding reverse link scenarios present in modern wireless infrastructures (i.e. cellular or \emph{ad hoc} networks) \cite{ref4}.

A great deal of attention has been given to the reduction of the Bit-Error-Rate (BER) probability and/or the outage probability, thus striving for the performance improvement and the robustness enhancement of OFCDM systems \cite{ref2}-\cite{ref1}. One of the most effective approaches is the subcarrier allocation adaptation, which is based on a subcarrier grouping methodology \cite{ref2}, according to the fading characteristics of each user and the channel load. In order to further improve this methodology, we propose a \emph{receiver-centric} approach (where the receiver rather than a potential sender controls the admission decision of a new transmission establishment), in order to exploit the diversity gain provided by the subcarrier grouping. It was shown in \cite{ref6} that such an approach significantly enhance the system performance of one dimensional (1D) spreading-enabled networks, in comparison to a conventional sender-centric one.

Nevertheless, the receiver-centric approach has not been studied for 2D spreading schemes so far. We propose a novel receiver-centric 2D OFCDM approach with subcarrier grouping and we manifest its performance enhancement in terms of the system robustness in comparison to the conventional 2D OFCDM approach. The advantages of the proposed scheme are three-fold: 1) Each sender-receiver active link does not need to have the knowledge of other ongoing transmissions within the same subcarrier group in terms of resource allocation and tolerable interference at the receiver; 2) It is fully distributed since the presence of a centralized unit, coordinating the user transmissions, is not required; 3) Accurate interference estimation can be achieved at the receiver, while a low information exchange overhead between the potential senders and the receiver is required.

\section{System Model}
\subsection{Transmission Establishment}
We consider an OFCDM system with a total number of $M$ subcarriers, which are separated into noncontiguous groups denoted as
\begin{equation}
G_{y}=\left\{m_{y}, m_{y+\mu}, m_{y+2\mu},..., m_{y+(M_{y}-1)\mu}\right\},
\end{equation}
where $y=\left\{1, 2,..., Y\right\}$, $\mu$ is the subcarrier spacing and $M_{y}$ is the number of subcarriers in group $y$. By applying a subcarrier grouping, an overall interference reduction caused by the frequency-domain spreading is achieved \cite{ref2}. The system supports up to maximum number of users, e.g. $\mathcal{B}$ users. Upon the transmission of the \textit{k}th user, each data bit experiences a 2D spreading by using a unique PseudoNoise (PN) code $C_{\mathcal{SF}}^{(k)}=M_{y}\times N$, where $N$ denotes the time-domain spreading length. The channel is assumed to be slowly varying with respect to the symbol duration $T_{b}$ and modeled as an independent Rayleigh fading for different users, subcarriers and bit transmissions.

In this letter, we adopt a receiver-centric OFCDM system by defining two successive user transmission phases, namely the \emph{probing} and the \emph{data phase}. The traffic arrives at each user in bursts of the same size. Prior to the actual data transmission, a potential sender enters the probing phase, by sending a signaling tone at the receiver. This tone carries no data information and actually represents a replica of a probing code, $C_{\mathcal{P}}$, common to all the system users. The signaling tone is used by the receiver to experience a fraction of the upcoming interference that a potential sender would cause and to calculate the influence that would produce (by entering the data phase) to the ongoing transmissions of other users. Furthermore, the statistics derived from the probing phase indicate each user's status at the receiver in terms of the signal fading and, thus, all the users can be appropriately allocated to the corresponding subcarrier group.

Since we assume independent Rayleigh fading coefficients for each user in the frequency-domain and a slowly varying channel within each OFCDM frame (i.e. the \textit{m}th fading entry of each user remains constant in the time-domain of the same OFCDM frame), we define the probing code as $C_{\mathcal{P}}=M_{y}\times N_{\mathcal{P}}$, where $N_{\mathcal{P}} \in (0, N]$ is unit-valued. Hence, $C_{\mathcal{P}}$ indicates to the receiver the previously mentioned amount of fading with a minimum interference cost, by applying a time-domain spreading gain which is only equal to one $C_{\mathcal{SF}}$ chip duration. Moreover, each system user is assigned a personal identifier, directly associated with a unique $M_{y}\times N_{\mathcal{P}}$ pair, to be utilized for the $C_{\mathcal{P}}$ transmission.

This results to the isolation of the signaling tone attempts, within the two dimensions of an OFCDM frame, for different users that enter to the probing phase. Consequently, the receiver identifies each user by simply locating its $C_{\mathcal{P}}$ position during an OFCDM frame reception. Hence, by using the proposed $C_{\mathcal{P}}$, an overall energy efficiency of the system is provided while a multiuser diversity gain is achieved. Note that by applying the proposed method, an occurrence of a collision event, caused by potentially simultaneous signaling tone transmissions from different users, is fully avoided. 

\begin{figure}[!t]
\centering
\includegraphics[keepaspectratio,width=2.8in]{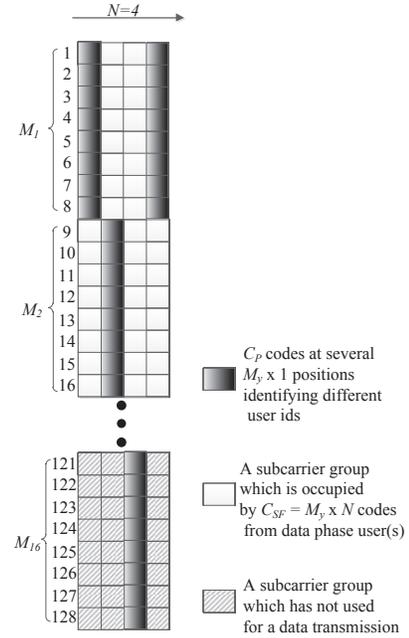}
\caption{A typical example of the proposed transmission approach with 128 subcarriers per OFCDM frame, when $C_{\mathcal{SF}}=8\times 4$.}
\label{Fig. 3}
\end{figure}

In the data phase, all the senders within a certain group utilize both the time and the frequency diversity of the channel, each by using its respective $C_{\mathcal{SF}}$, which is \emph{a priori} known at the receiver. Thereupon, a potential sender keeps transmitting partial chunks of the remaining burst at all the subsequent OFCDM frames, on its dedicated $G_{y}$, until the entire burst is completed. Fig. ~\ref{Fig. 3} shows a typical example of the proposed OFCDM 2D spreading approach, when $C_{\mathcal{SF}}=8\times 4$.

\subsection{Reception Strategy}
The overall received signal is expressed as
\begin{align}
\nonumber
r(t)&=\sqrt{2\epsilon_{c}}\bigg\{\sum_{k=1}^{K}\sum_{j}b^{k}_{j}\sum_{m=1}^{M}v^{k}_{j,m}\alpha^{k}_{j,m}c^{F(k)}_{j,m}\Omega^{k}_{j,m}\\
&\times \sum_{n=1}^{N} c^{T(k)}_{j,m,n} p[t-(jN+n)T_{c}]\\
\nonumber
&+ \sum_{k'=1}^{K'}\sum_{m=1}^{M}v^{k'}_{j,m}\alpha^{k'}_{j,m}c^{F(k')}_{j,m}\Omega^{k'}_{j,m}\\
\nonumber
&\times \sum_{n=1}^{N} c^{T(k')}_{j,m,n} p[t-(jN+1)T_{c}]\bigg\}+n(t),
\end{align}
where $(.)'$, $T_{c}$, $\epsilon_{c}$ and $p[.]$ denote a probing instance, the chip duration, the chip energy and a rectangular pulse shaping filter, respectively. Assuming Binary Phase Shift Keying (BPSK) modulation schemes, $b^{k}_{j}=\pm 1, j \in \aleph$. The term $\alpha^{k}_{j,m}$ is the Rayleigh fading gain for the \textit{m}th subcarrier of the \textit{k}th user, during the \textit{j}th transmitted bit. The term $\Omega^{k}_{j,m}=cos(\omega_{m}t+\phi^{(k)}_{j,m})$, while $\omega_{m}$ and $\phi^{(k)}_{j,m}$ denote the \textit{m}th subcarrier frequency and the respective phase shift (produced by the channel) of the received signal for the \textit{k}th user. The frequency-domain and the time-domain PN chips are, respectively, expressed as $c^{F(.)}$ and $c^{T(.)}$. The parameter $v^{k}_{j,m}$ equals one $iff$ $m \in G_{y}$ and zero otherwise. Finally, $n(t)$ is the Additive White Gaussian Noise (AWGN) signal with a double-sided power spectral density of $N_{\textit{o}}/2$.

In order to accurately decode the data of a user, e.g. user 1, the received signal is restored to the baseband by multiplying it with $\sqrt{2}cos(\omega_{m}t-\phi^{(1)}_{j,m})$, followed by a low-pass filtering. Moreover, the interference term of user 1 is obtained as
\begin{align}
\nonumber
&I^{(1)}_{j}=\sqrt{\epsilon_{c}}\Big\{\sum_{k \in G_{y}}b^{(k)}_{j}\sum_{m \in G_{y}}\alpha^{(1)}_{j,m}\alpha^{(k)}_{j,m}c^{F(1)}_{j,m}c^{F(k)}_{j,m}\Delta_{\phi}^{k,1}\\
\nonumber
&\times\frac{1}{T_{c}}\int^{(j+1)T_{b}}_{t=jT_{b}}\sum^{N}_{n=1}c^{T(1)}_{j,m,n}c^{T(k)}_{j,m,n}p[t-(jN+n)T_{c}]dt\\
&+\sum_{k' \in G_{y}}\sum_{m \in G_{y}}\alpha^{(1)}_{j,m}\alpha^{(k')}_{j,m}c^{F(1)}_{j,m}c^{F(k')}_{j,m}\Delta_{\phi}^{k',1}\\
\nonumber
&\times\frac{1}{T_{c}}\int^{(j+\frac{1}{N})T_{b}}_{t=jT_{b}}\sum^{N}_{n=1}c^{T(1)}_{j,m,n}c^{T(k')}_{j,m,n}p[t-(jN+1)T_{c}]dt\Big\},
\end{align}
where $\Delta_{\phi}^{k,1}=\phi^{(k)}_{j,m}-\phi^{(1)}_{j,m}$.

Since the bit information $b_{j}^{(1)}$ has zero mean, the power of the desired signal is derived by simply calculating its variance as 
\begin{equation}
P^{(1)}_{d_{j}}=N^{2}\epsilon_{c}\left(A^{(1)}_{j,m}\right)^{2},
\end{equation}
where $A^{(1)}_{j,m}=\sum_{m\in G_{y}}(\alpha^{(1)}_{j,m})^{2}$. Finally, the noise power is expressed as
\begin{equation}
P^{(1)}_{n_{j}}=NN_{\textit{o}}\sum_{m\in G_{y}}\left(\alpha^{(1)}_{j,m}\right)^{2}.
\end{equation}

By using the \emph{Central Limit Theorem}, the power of $I^{(1)}_{j}$ can be approximated as a Gaussian random variable, which produces good results for a sufficiently large number of users and is obtained as \cite{ref1,ref4}
\begin{align}
P^{(1)}_{i_{j}}&= \sum_{m\in G_{y}}(\alpha^{(1)}_{j,m})^{2}\\
\nonumber
&\times\left\{N\epsilon_{c}(K_{y}-1)\textit{E}\left[(\alpha_{y})^{2}\right]+\sum^{N}_{n=1}\epsilon_{c}v_{n}\textit{E}\left[(\alpha_{y})^{2}\right]\right\},
\end{align}
where
\begin{equation}
\textit{E}\left[(\alpha_{y})^{2}\right]=\frac{1}{\mathcal{B}_{y}M_{y}}\sum_{\beta\in \mathcal{B}_{y}}\sum_{m\in G_{y}}(\alpha^{(\beta)}_{y})^{2},
\end{equation}
and $\mathcal{B}_{y}= K'_{G_{y}} \cup K_{G_{y}}$, $K'_{G_{y}} \cap K_{G_{y}}=\emptyset$. The terms $K'_{G_{y}}$ and $K_{G_{y}}$ denote the total number of users within $G_{y}$ occurring in the probing phase and the data phase, respectively. Hence, the Signal-to-Interference-plus-Noise Ratio (SINR) of the \textit{k}th user on $G_{y}$, during the \textit{j}th bit transmission, is expressed as
\begin{align}
\nonumber
&\gamma^{(k)}_{j,G_{y}}=\\
&\frac{N^{2}\epsilon_{c}\sum_{m\in G_{y}}\left(\alpha^{(k)}_{m}\right)^{2}}{N\epsilon_{c}(K_{y}-1)\textit{E}\left[(\alpha_{y})^{2}\right]+\sum^{N}_{n=1}\epsilon_{c}v_{n}\textit{E}\left[(\alpha_{y})^{2}\right]+NN_{\textit{o}}},
\label{4}
\end{align}
where $v_{n}$ equals one $iff$ a signaling tone occurs within the \textit{n}th time instance of $G_{y}$ and zero otherwise.

\subsection{The Proposed Algorithm}
Our main objective is to maximize $\gamma^{(k)}_{j,G_{y}}$. The proposed algorithm is analytically presented in the following steps:

\begin{description}
	\item[\emph{Step 1}:] \hspace{0.1cm} During an OFCDM frame reception, the receiver checks if there is a probing tone $\forall n \in (0, N]$ within each $G_{y}$, $y \in (0, Y]$ and marks a potential probing event as $G_{y}(n)$, identifying the respective $k'$th user at the same time. 
\end{description}

\begin{description}
	\item[\emph{Step 2}:] \hspace{0.1cm} Based on eq. (\ref{4}), the receiver calculates $\gamma^{(k)'}_{G_{y}}$, $\forall y, k'$, which is expressed as
\end{description}
\begin{equation}
\gamma^{(k)'}_{G_{y}}=\\
\frac{N^{2}\epsilon_{c}\sum_{m\in G_{y}}\left(\alpha^{(k)'}_{m}\right)^{2}}{N\epsilon_{c}K_{y}\textit{E}\left[(\alpha_{y})^{2}\right]+\sum^{N}_{n=1}\epsilon_{c}v^{(k)'}_{n}\textit{E}\left[(\alpha_{y})^{2}\right]+NN_{\textit{o}}},
\end{equation}
\begin{description}
\item[]
where $v^{(k)'}_{n}$ equals one $iff$ a signaling tone occurs within the \textit{n}th time instance of $G_{y}$ (while excepting the under examination $k'$th one, i.e. $v^{(k)'}_{n}=1$, $\forall (.)'\neq (k)'$) and zero otherwise. Afterwards, the receiver finds $\gamma^{(k)'}_{G_{max}}=max[\gamma^{(k)'}_{G_{1}}, \gamma^{(k)'}_{G_{2}},..., \gamma^{(k)'}_{G_{Y}}]$, where $max[.]$ denotes the maximum operation.
\end{description}

\begin{description}
	\item[\emph{Step 3}:] \hspace{0.1cm} In the next OFCDM frame transmission duration, the receiver indicates $G^{(k)'}_{max}$ to the $k'$th user, which represents the most effective subcarrier group for this user (and, hence, for all the ongoing transmitting users that exploit the data phase), by sending a probing tone to $G_{max}(n)$.
\end{description}

\begin{description}
	\item[\emph{Step 4}:] \hspace{0.1cm} Thereupon, the $k'$th user initiates the data phase, by sending packets within $G_{max}$ using its personal $C_{\mathcal{SF}}$ at all the subsequent OFCDM frames, until its burst is completed. Note that the receiver \textit{a priori} knows the latter spreading code and since it expects to collect information from that user at $G_{max}$, the decoding is accurately performed.
\end{description}

Given the number of the consecutive OFCDM frames that are required for the completion of the data phase for each user, denoted as $\xi$, the transmission duration of the entire burst can be expressed as $\xi \times M T_{b} (=\xi \times M N T_{c})$. Consequently, by adopting the proposed probing approach, the corresponding transmission duration of a given burst is extended to $(\xi+2) \times M T_{b}$. The extra 2 OFCDM frames represent an outcome of the probing methodology, i.e. one for the probing transmission/reception (\emph{Step 1}) and one for the optimization procedure (\emph{Step 3}). It is worth noting that for large values of $\xi$, which is usually the case in realistic network implementations, the cost that is produced by the probing overhead can assumed to be negligible.

\section{Performance Evaluation}
\subsection{BER Performance}
\begin{figure}[!t]
\centering
\includegraphics[keepaspectratio,width=3.4in]{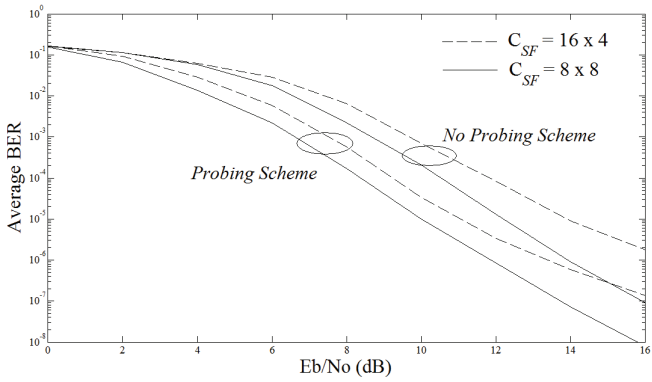}
\caption{BER vs. $Eb/N_{\textit{o}}$, for 32 users and different spreading factors.}
\label{Fig. 1}
\end{figure}

The BER performance of the proposed approach is evaluated by using Monte Carlo simulations. The OFCDM system bandwidth is assumed to be 20 MHz, consisting of 128 subcarriers. The system supports up to 32 users, the burst arrival at each user is a Poisson process, while the burst size is $10^{6}$ bits. Since we focus on BPSK constellation alphabets, the BER probability for the \textit{k}th user, is expressed as 
\begin{equation}
Pe^{(k)}\triangleq Q\left(\sqrt{2\overline{\gamma}^{(k)}_{G_{y}}}\right),
\end{equation}
where $\overline{\gamma}^{(k)}_{G_{y}}$ is obtained by averaging over $\gamma^{(k)}_{j,G_{y}}$, while $Q(.)$ is the Gaussian \textit{Q}-function. In Fig.~\ref{Fig. 1}, along with the proposed \textit{probing} scheme we also modeled the classical OFCDM receiver (such as \cite{ref1} and \cite{ref4}), denoted as \textit{no probing} scheme, for a cross-reference comparison. As an illustrative example, we consider two case studies, each with a different spreading gain, namely $C_{\mathcal{SF}}=16\times 4$ and $C_{\mathcal{SF}}=8\times 8$. It is obvious that the latter spreading gain results to a superior BER performance for both schemes in comparison to the former one, due to the higher frequency and time diversity gain, since we apply a subcarrier grouping policy. The proposed scheme outperforms the conventional OFCDM scheme, as shown in Fig.~\ref{Fig. 1} (i.e. an improvement of approximately 2.5 dB gain is achieved), which is an outcome of the proposed probing method. 

Note that the above-mentioned performance gain is directly associated with the structure of $C_{\mathcal{P}}$. The design of the \emph{column-wise} $C_{\mathcal{P}}$ instances, as explicitly defined in the previous section, counteracts the fragility between the multiuser diversity and the error resilience of the system, which is suitably effective under slowly varying fading channels. Nevertheless, the structure of $C_{\mathcal{P}}$ has to be appropriately re-defined in case of severe fading conditions (e.g. when doubly selective channel fading is present). Such a probing approach should compass for both frequency and time selectivity of the channel (conditioned on the respective coherence time) while maintaining the multiuser diversity gain in an acceptable level and it represents a challenging topic for further research.

\subsection{Outage Probability}
\begin{figure}[!t]
\centering
\includegraphics[keepaspectratio,width=3.4in]{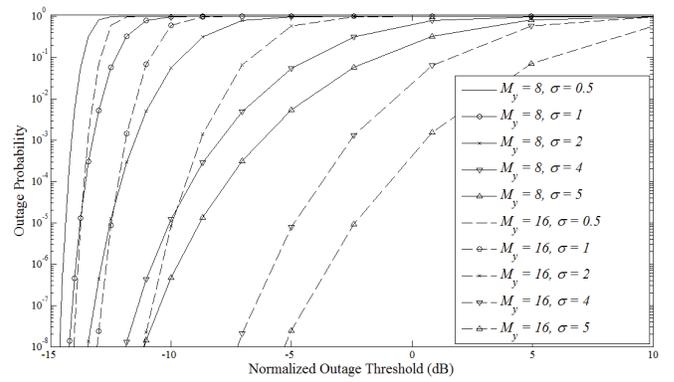}
\caption{$P_{out}$ vs. $\gamma_{th}/\overline{\gamma}_{G_{y}}$, for several values of $M_{y}$ and $\sigma$, based on the closed-form approximation given in eq. (\ref{6}).}
\label{Fig. 2}
\end{figure}

The outage probability denotes the probability given that the instantaneous SINR falls below a specified threshold, $\gamma_{th}$, and can be obtained as \cite{ref5}
\begin{equation}
P_{out}\triangleq Pr(\gamma \leq \gamma_{th})=\int^{\gamma_{th}}_{0}f_{\gamma}(\gamma)d\gamma=\mathcal{F}_{\gamma}(\gamma_{th}),
\end{equation}
where $f_{\gamma}(.)$ and $\mathcal{F}_{\gamma}(.)$ denote, respectively, the probability density function (PDF) and the cumulative density function (CDF) of the instantaneous SINR. Since $\gamma(.)$ is directly related to the sum of i.i.d. Rayleigh parameters, ranging according to the $M_{y}$ size, we use a highly accurate closed-form approximation \cite{ref3} to estimate $\mathcal{F}_{\gamma}(\gamma_{th})$, which is expressed as
\begin{align}
\nonumber
\mathcal{F}_{\gamma}(\gamma_{th})&\cong 1-\textit{e}^{-\frac{\gamma_{th}^{2}}{2\psi}}\sum_{m=0}^{M_{y}-1}\frac{\left(\frac{\gamma_{th}^{2}}{2\psi}\right)^{m}}{m!}\\
&-\gamma_{th}\frac{\alpha_{0}(\gamma_{th}-\alpha_{2})^{2M_{y}-1}\textit{e}^{-\frac{\alpha_{1}(\gamma_{th}-\alpha_{2})^{2}}{2\psi}}}{2^{M_{y}-1}\left(\frac{\psi}{\alpha_{1}}\right)^{M_{y}}(M_{y}-1)!},
\label{6}
\end{align}
where $\psi=\frac{\sigma}{M_{y}}[(2M_{y}-1)!!]^{1/M_{y}}$, $\sigma$ is the distribution shaping parameter and $(2M_{y}-1)!!=(2M_{y}-1)(2M_{y}-3)\cdot\cdot\cdot3\cdot1$. The constants $\alpha_{0}$, $\alpha_{1}$ and $\alpha_{2}$ are used to enhance the estimation accuracy of the CDF approximation and depend on the length of $M_{y}$. The particular values for these constants, utilized in this letter, are presented in Table ~\ref{Table I} (a detailed representation of the range of $\alpha_{0}$, $\alpha_{1}$ and $\alpha_{2}$ can be found in \cite[Table I]{ref3}). As shown in Fig.~\ref{Fig. 2}, the outage performance improves (i.e. $P_{out}$ decreases) more sharply with an increase of $M_{y}$, especially at the low SINR regions, with respect to the normalized outage threshold ($\gamma_{th}/\overline{\gamma}_{G_{y}}$). This is achieved due to the frequency-domain diversity gain, as the size of $M_{y}$ increases, at the cost of the multiuser diversity loss and vice versa.
\begin{table}[!ht]
\centering
\caption{Constant Values Used in the Approximation of eq. (\ref{6})}
\label{Table I}
\begin{tabular}
{ c| c c c }
\hline
$M_{y}$ size & $\alpha_{0}$ & $\alpha_{1}$ & $\alpha_{2}$ \\
\hline
8 & 0.0257 & 0.1172 & 0.9491 \\
16 & 0.0291 & 0.0133 & 0.9338 \\
\hline
\end{tabular}
\end{table}

\section{Conclusions}
A novel data transmission approach for multiuser 2D spreading OFCDM systems is presented. The receiver, rather than a potential sender, decides for the admission of a data transmission between each sender-receiver pair. A subcarrier grouping policy is established and the assignment of each user to the system is based on the channel load and the SINR status at each subcarrier group. This is accomplished by addressing two transmission phases for the potential system users, namely the probing and the data phase. We showed that the proposed approach outperforms the conventional OFCDM scheme with respect to the system robustness and the error resilience.

\ifCLASSOPTIONcaptionsoff
  \newpage
\fi

\end{document}